\documentclass[aps,prx,twocolumn,superscriptaddress,longbibliography,nobibnotes]{revtex4-2}
\usepackage{diagbox}
\usepackage{amsmath}
\usepackage{amsfonts}
\usepackage{physics}
\usepackage[space]{grffile}
\usepackage{graphicx} 
\usepackage{amssymb}
\usepackage{upgreek}
\usepackage{textgreek}
\usepackage{lineno}
\usepackage{natbib}
\usepackage{braket}
\usepackage{float}
\usepackage{color}
\usepackage{blindtext}
\usepackage{comment}
\usepackage{bm}
\setcitestyle{super} 

\usepackage{xr-hyper}
\usepackage{hyperref}
\usepackage{xcolor}
\hypersetup{colorlinks,breaklinks,
            urlcolor=[rgb]{0,0,0.64},
            linkcolor=[rgb]{0,0,0.64},
            citecolor=[rgb]{0,0,0.64},
            filecolor=[rgb]{0,0,0.64}}

\usepackage[T1]{fontenc}

\makeatletter
\renewcommand{\thefigure}{\textbf{\@arabic\c@figure}}
\makeatother

\newcommand{\ColumbiaPhys}{Department of Physics, Columbia University, New York, NY, USA.}
\newcommand{\ColumbiaChem}{Department of Chemistry, Columbia University, New York, NY, USA.}
\newcommand{\ColumbiaME}{ Department of Mechanical Engineering, Columbia University,New York, NY, USA.}
\newcommand{\NIMSichi}{ Research Center for Materials Nanoarchitectonics, National Institute for Materials Science, Tsukuba, Japan.}
\newcommand{\NIMSni}{ Research Center for Electronic and Optical Materials, National Institute for Materials Science, Tsukuba, Japan.}
\newcommand{\MPSD}{Max Planck Institute for the Structure and Dynamics of Matter, Hamburg, Germany.}
\newcommand{\UdPV}{Nano-Bio Spectroscopy Group, Departamento de Fisica de Materiales, Universidad del Pais Vasco, San Sebastian, Spain.}
\newcommand{\FlatIron}{Initiative for Computational Catalysis (ICC) and Center for Computational Quantum Physics (CCQ), Flatiron Institute, New York, New York, USA.}

\makeatletter

\let\rtx@note\relax
\def\@author@join@#1#2#3{%
  \def\@author{{#1}{#2\rtx@note{#3}}}%
}

\def\rtx@note@first#1{%
  \frontmatter@footnote{#1}%
  \let\rtx@note\rtx@note@next
}

\def\rtx@note@next#1{%
  \comma@space
  \frontmatter@footnote{#1}%
}

\def\doauthor#1#2#3{%
  \ignorespaces#1\unskip\@listcomma
  \begingroup
    \def\rtx@tmp{#2}%
    \ifx\rtx@tmp\@empty
      \def\rtx@finish{\endgroup{}{}}%
    \else
      \def\rtx@finish{%
        \endgroup{\comma@space}{}%
        \let\rtx@note\rtx@note@first
        #2%
      }%
    \fi
    #3%
  \rtx@finish
  \space\@listand
}

\makeatother

\begin{document}

\title{Antiferromagnetism-altered plasmon dynamics}


\author{Yifan~Su}
\thanks{These authors contributed equally: Y.S. and S.X.}
\email[Correspondence to: ]{ys3945@columbia.edu}
\affiliation{\ColumbiaPhys}
\author{Suheng~Xu}
\thanks{These authors contributed equally: Y.S. and S.X.}
\author{Rocco~A.~Vitalone}
\affiliation{\ColumbiaPhys}
\author{Ziyu~Liu}
\affiliation{\ColumbiaPhys}
\author{Emil~Vi\~{n}as~Bostr\"{o}m}
\affiliation{\MPSD}
\author{Na~Wu}
\affiliation{\MPSD}
\author{Chun-Ying Huang}
\affiliation{\ColumbiaChem}
\author{Zhuquan~Zhang}
\affiliation{\ColumbiaPhys}
\affiliation{\ColumbiaChem}
\author{Jaehoon~Jung}
\affiliation{\ColumbiaPhys}
\author{Daniel~G.~Chica}
\affiliation{\ColumbiaChem}
\author{Vin\'{i}cius~da~Silveira~Lan~Avelar}
\affiliation{\ColumbiaChem}
\author{Yiping~Wang}
\affiliation{\ColumbiaChem}
\affiliation{\ColumbiaME}
\author{Takashi~Taniguchi}
\affiliation{\NIMSichi}
\author{Kenji~Watanabe}
\affiliation{\NIMSni}
\author{James~C.~Hone}
\affiliation{\ColumbiaME}
\author{Xiaoyang~Zhu}
\affiliation{\ColumbiaChem}
\author{Cory~R.~Dean}
\affiliation{\ColumbiaPhys}
\author{Xavier~Roy}
\affiliation{\ColumbiaChem}
\author{Angel~Rubio}
\affiliation{\MPSD}
\affiliation{\UdPV}
\affiliation{\FlatIron}
\author{D.~N.~Basov}
\email[Correspondence to: ]{db3056@columbia.edu}
\affiliation{\ColumbiaPhys}

\date{\today}

\begin{abstract}
The interaction between plasmons and magnons is a long-sought phenomenon with implications for fundamental physics and spintronics applications. In three-dimensional systems, this coupling is suppressed by the large mismatch in energy scales, but two-dimensional (2D) plasmons with gapless dispersion can overlap with magnons over a broad spectral range. Despite numerous theoretical predictions, experimental observation of magnon-plasmon interaction has remained elusive. In this work, we study a first-of-its-kind hybrid plasmon-magnon platform based on 2D materials. By deploying scattering-type scanning near-field optical microscopy (s-SNOM) with terahertz radiation, we image propagating plasmon wavepackets at a graphene/NiPS$_3$ interface and track their dynamics across the antiferromagnetic transition of NiPS$_3$. We observe a clear renormalization of the plasmon-polariton dispersion concurrent with the onset of antiferromagnetic order. With complementary Raman scattering and nano-terahertz spectroscopy, we unveil spectral weight redistribution and dielectric screening changes, potentially associated with the multi-magnon continuum, as the underlying mechanism. These results provide solid evidence of coupling between plasmon and antiferromagnetic order, marking a cornerstone for a potential platform for hybrid magnon–plasmon interactions in 2D materials, opening avenues for coherent spin–plasmon devices and tunable terahertz spintronic components.
\end{abstract}

\maketitle

\section{Introduction}

\begin{figure}[htb!]
\centering
\includegraphics[width=0.45\textwidth]{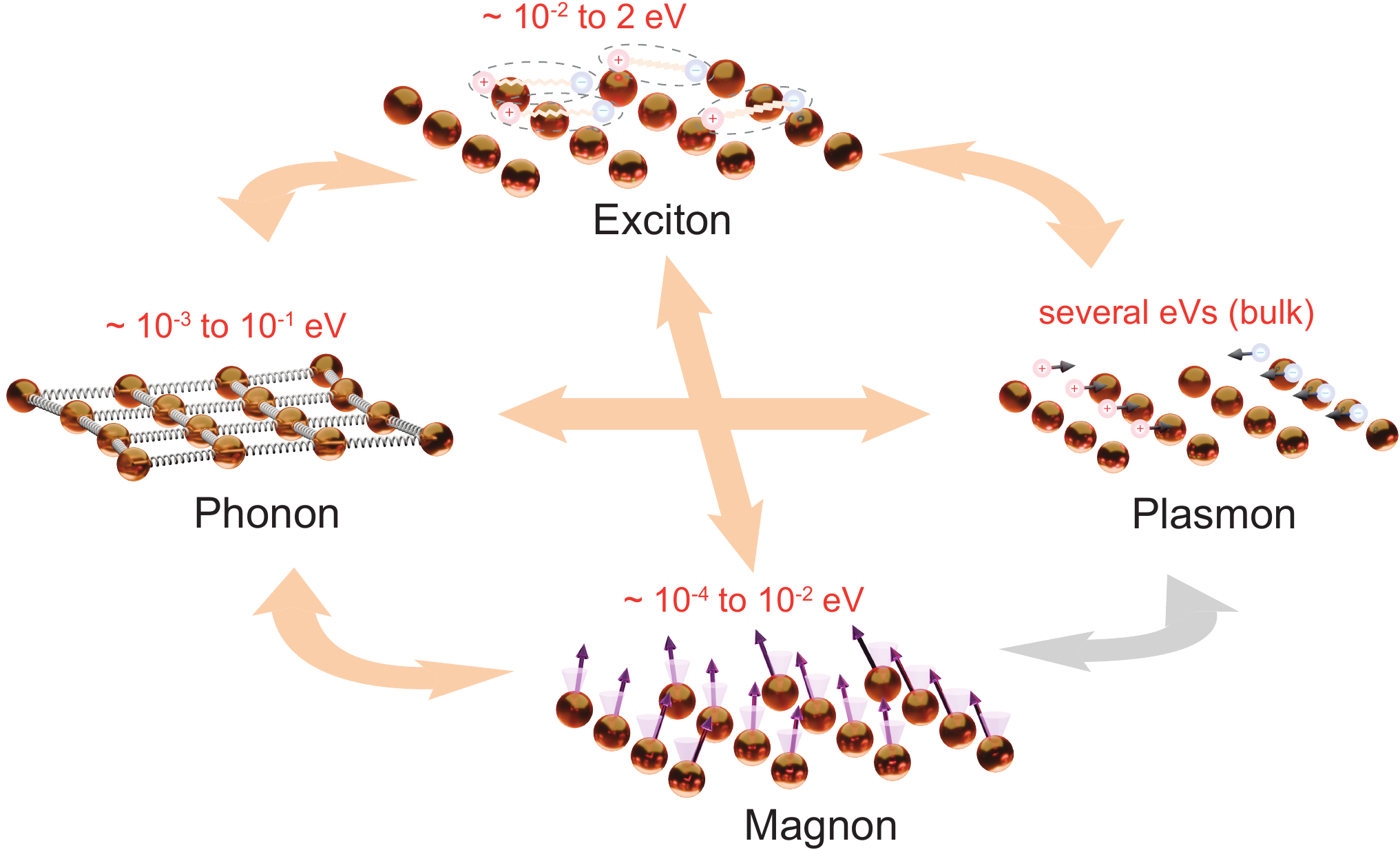}
\caption{\textbf{Typical fundamental excitations in solids and the corresponding energy scales.} Four common types of bosonic collective excitations that are intensely studied in condensed matter physics and correspondingly their typical energy scales. Among each pair of excitations, orange arrows in between indicate previously observed interactions, while the gray arrow indicates the previously unseen interaction. }
\label{fig:Fig0}
\end{figure}

The interactions among bosonic excitations in quantum materials systems remain the central topic in condensed matter physics. The last decade witnessed a boom of pairwise interactions, as indicated in Fig.~\ref{fig:Fig0}(a), among the four most common types of excitations --- phonons, magnons, excitons, and plasmons. Plasmon-phonon interaction can be routinely observed in graphene-hBN interfaces as hybridized phonon-plasmon-polaritons\cite{Brar2014}. Similarly, excitons and plasmons can hybridize into exciton-plasmon-polaritons in a gold-semiconductor interface
\cite{zhang2020,Moore2025}. Excitons and phonons can interact naturally through the band structure change induced by lattice distortion, and the interaction has been observed in various semiconductor systems\cite{Ning2020,Chen2025}. Recent advances in van der Waals magnets provided a plethora of magnon-boson interaction platforms\cite{Park2025}. Van der Waals antiferromagnets NiPS$_3$ and CrSBr feature spin-orbit-entangled excitons that are strongly coupled to antiferromagnetic magnons\cite{Kang2020,Belvin2021,Bae2022}. FePS$_3$ and FePSe$_3$, on the other hand, feature significant magnetostriction and magnetoelasticity effects that lead to strong phonon-magnon coupling and allow for coherent ultrafast control of magnetism through the phonon structure, and vice versa\cite{Zong2023,Zhou2022,Ilyas2024,Luo2025}. Nonetheless, despite the numerous observations of various interactions among different collective excitations, clear experimental observations of magnon-plasmon interaction remain absent. 

The elusive magnon-plasmon interaction is not merely a lost piece in fundamental physics sought for pure intellectual curiosity, but also a missing yet crucial element that bridges two fields in applied science and engineering: spintronics and plasmonics\cite {Zutic2004,Lin2019,Fei2012,Grigorenko2012,Ni2018,Xu2024}. 
By integrating spintronics and plasmonics, emergent magnetoplasmonics, or spinplasmonics, promises to combine the information‑rich, coherent spin degrees of freedom in spintronics with ultra-confined, electrically reconfigurable electromagnetic modes in plasmonics. 
Such coupling could enable nanoscale spin–charge transduction and electrically tunable hybrid spin–electromagnetic modes, with potential applications in terahertz and quantum technologies. \cite{Basov2016,Basov2025,Elezzabi2008,Maksymov2016,Singh2023}. 
The two-fold significance of magnon-plasmon interaction in both fundamental and applied science thus spurs the interest in searching for this missing piece of the puzzle.

Historically, one of the major reasons behind the scarcity of magnon-plasmon interaction was their vastly different energy scales. As shown in Fig.~\ref{fig:Fig0}, magnons, the quantized spin-wobbling modes in solids, typically occur at energies of about several millielectronvolts for antiferromagnetic (AFM) systems and hundreds of microelectronvolts for ferromagnetic (FM) systems. This intrinsic energy scale is determined by the nature of spin-spin interactions. On the other hand, in three-dimensional (3D) metallic systems, plasmons, the quantized charge oscillation modes, have a typical energy scale of several or tens of electronvolts, governed by the energy scale of Coulomb interactions. This approximately 4-6 orders of magnitude difference in energy scales makes it nearly impossible to realize magnon-plasmon interactions in the bulk of 3D systems.

Meanwhile, the physical rules in two-dimensional (2D) systems tell a different story. Plasmons in 2D systems hybridize with electromagnetic fields and form surface plasmon-polaritons (SPPs), taking the form of confined hybrid light-matter modes with gapless dispersion\cite{Ni2018,Basov2016,Basov2025}. The gapless dispersion and electromagnetic wave components make it possible for SPPs to interact with magnons, both in energy scales and fundamental interactions. 

\begin{figure*}[htb!]
\centering
\includegraphics[width=0.9\textwidth]{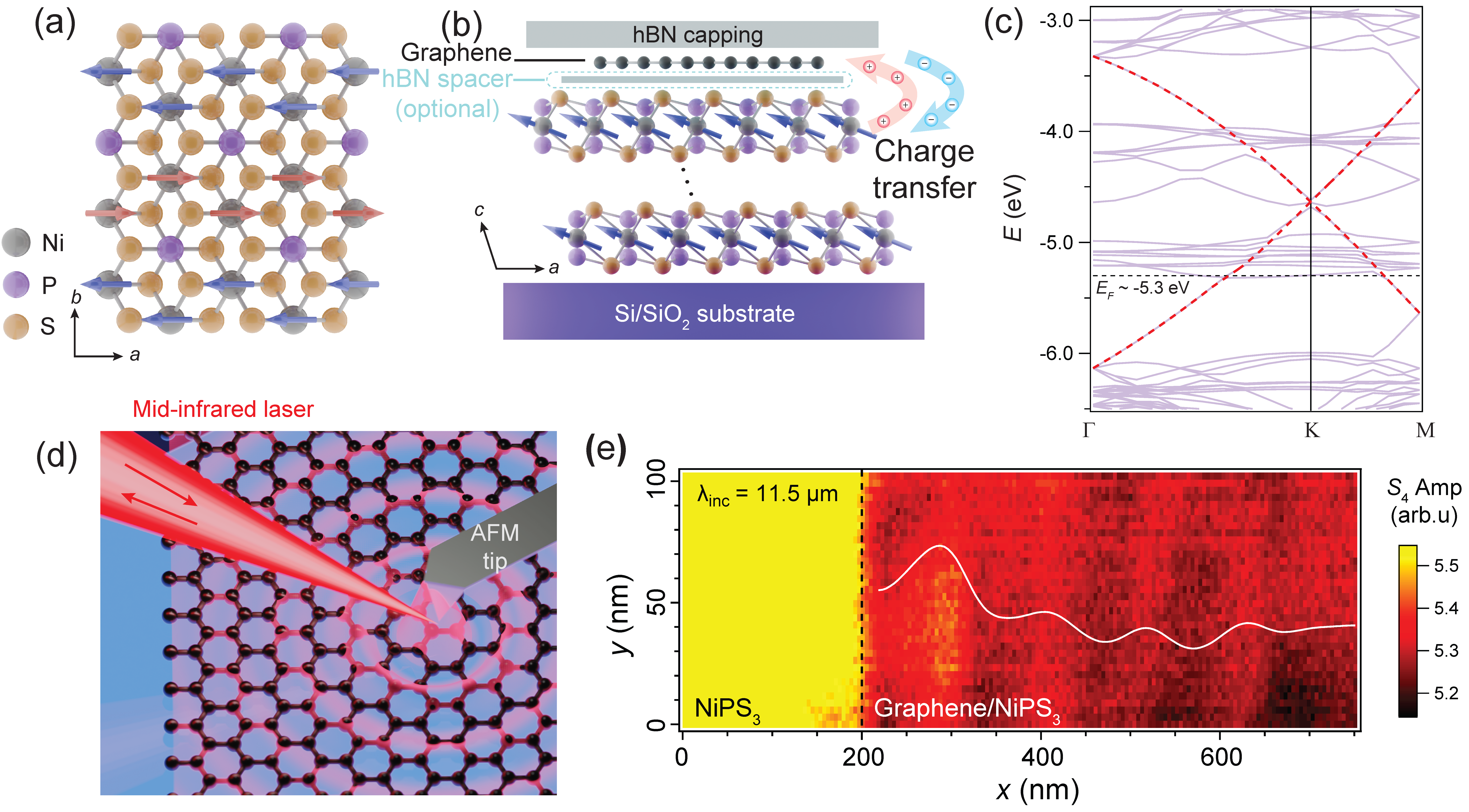}
\caption{\textbf{Magneto-plasmonics at the Graphene/NiPS$_3$ interface via charge transfer.} (a-b) Crystal and magnetic structure of NiPS$_3$, and device geometry of Graphene/NiPS$_3$ device. (a)~In-plane crystal and magnetic structure of NiPS$_3$. The Ni atoms (grey spheres) are arranged in a two-dimensional honeycomb lattice in the $ab$-plane. The red and blue arrows indicate opposite spin orientations in the zig-zag antiferromagnetic order. (b) The out-of-plane structure of NiPS$_3$ and device geometry. The monoclinically stacked NiPS$_3$ layers are coupled by a weak ferromagnetic interaction. Electrons in graphene transfer to the top few layers in NiPS$_3$ when put in contact with NiPS$_3$ due to the different work functions. An optional hBN spacer layer can be deposited between NiPS$_3$ and graphene for tuning charge transfer and improving the spatial homogeneity of the interface. (c) Band structure of graphene/NiPS$_3$ interface (with hBN spacer) from density functional theory. Dashed red lines highlight the graphene band, featuring a hole-doped Dirac cone due to interfacial charge transfer. (d) Schematics of scattering-type Scanning Near-field Optical Microscopy (s-SNOM) in the mid-infrared (MIR) regime in backward scattering geometry. A MIR beam from a quantum cascade laser illuminates a metal-coated atomic force microscope (AFM) tip and generates surface plasmon-polaritons (SPPs). (e) s-SNOM image of a graphene/NiPS$_3$ interface with hBN spacer, with an incident MIR laser wavelength of 11.5~$\upmu$m, showing oscillations in the near-field $S_4$ amplitude that are characteristic of SPPs. The black dashed line denotes the boundary of the graphene. The white solid curve denotes the integrated intensity of $S_4$ in graphene along the edge (vertical) direction.}
\label{fig:Fig1}
\end{figure*}

In the past several years, multiple theoretical works predicted the possible realization of magnon-plasmon interactions in 2D systems\cite{Costa2023,Yuan2025,Gorobets2025,Dyrda2023}, while the experimental realization of magnon-plasmon interactions still remains unexplored. Though magnon-plasmon interaction is theoretically predicted in 2D systems\cite{Costa2023,Yuan2025,Gorobets2025,Dyrda2023}, there are several factors that account for the experimental obscurity. First and foremost, magnetic-dipole transitions are intrinsically weak compared with electric-dipole processes, resulting in weak magnon-plasmon coupling and making the interaction difficult to observe experimentally. In addition, atomically thin magnets are often fragile with small optical cross sections and require precise heterostructure assembly. Moreover, the need for simultaneous control of electronic doping, magnetic ordering, and near‑field coupling significantly constrains the viability for the experimental observation of magnon-plasmon coupling.

While these constraints impose significant challenges to the experimental realization of magnon-plasmon interaction, they, at the same time, set a clear road-map that guide the design for a hybrid magnon-plasmon platform: exploit low‑loss 2D plasmonic media and robust 2D magnets, optimize near‑field engineering for mode overlap, and control doping to achieve the coherent strong‑coupling regime that would unlock new fundamental physics and device functionalities at the crossroads of spintronics and plasmonics.

\section{Results}

In this work, we present a van der Waals heterostructure platform that is ideally suited for studying the interaction between magnetic order and 2D plasmons, probed by a state-of-the-art scanning near-field terahertz nanoscopy tool that enables direct access to plasmonic waves in hybrid 2D magneto-plasmonic systems. With the novel material system and experimental techniques, we demonstrate a novel phenomenon of tuning 2D plasmonics with magnetic order.

\begin{figure*}[htb!]
\centering
\includegraphics[width=\textwidth]{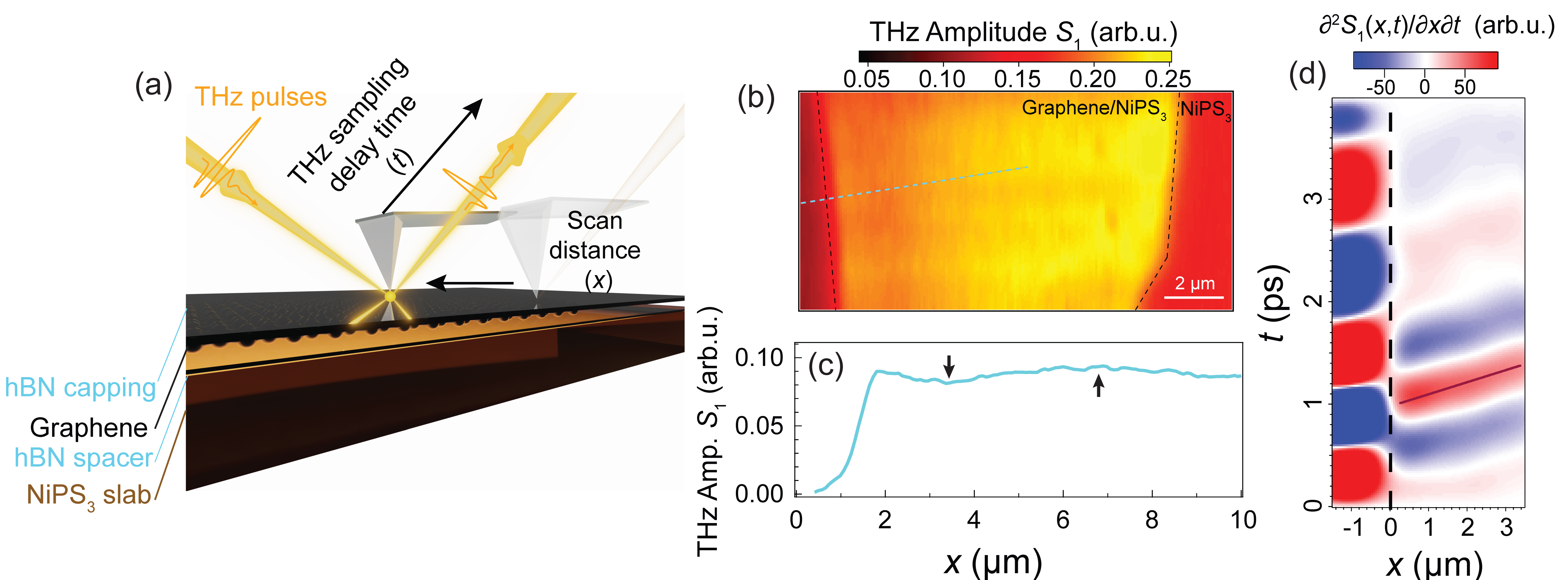}
\caption{\textbf{Terahertz imaging and spacetime mapping for measuring SPP propagation.} (a) Schematic illustration of terahertz (THz) s-SNOM imaging and spacetime mapping experiments. The graphene/NiPS$_3$ stack is illuminated with broadband THz pulses (0.5 to 1.5 THz). The tip acts as a nano-antenna, launching and detecting SPP wavepackets, whose propagation dynamics can be recorded by scanning the tip position ($x$) and THz sampling delay time ($t$) (b) Nano-THz amplitude image of monolayer graphene on top of NiPS$_3$, separated by hBN spacer layers. The cyan dashed line indicates the scanning path for the spacetime map measurement in (c-d). (c) Amplitude line profile along the cyan dashed line in (b). The dip and hump structure indicated by the black arrows signatures the plasmon-polariton in graphene. (d) Spacetime map of the near-field plasmon electric field at 300~K. The measurement is performed by recording the THz scattering amplitude ($S_1$) along the cyan dashed line in (b) at different delay time points of the THz pulse sampling. The background THz field is removed by taking the partial derivatives with respect to both spatial ($x$) and time ($t$) axes, so that the plasmonic signal is highlighted. The black dashed line denotes the graphene boundary.}
\label{fig:Fig2}
\end{figure*}

The materials platform we designed consists of two subsystems: a magnetically-order material, NiPS$_3$, and a plasmonic medium, monolayer graphene. NiPS$_3$ is a van der Waals antiferromagnet, in which Ni atoms are arranged in a 2D honeycomb lattice and their magnetic moments form a zigzag antiferromagnetic order below the N\'{e}el temperature ($T_N \approx$  155 K)\cite{Kang2020,Belvin2021}, as shown in Fig.~\ref{fig:Fig1}\textbf(a,b). The design of the magneto-plasmonic hetero-structure is simple: we stack a graphene monolayer on top of a thick microcrystal of NiPS$_3$, with an optional hexagonal boron nitride (hBN) spacer in between, as demonstrated in Fig.~\ref{fig:Fig1}(b). Monolayer graphene proves to support a high-quality 2D plasmon mode\cite{Fei2012, Basov2016,Ni2018,Basov2025} when doped. The NiPS$_3$ microcrystal in this structure serves a dual purpose. Besides the antiferromagnetic order, the larger work function of NiPS$_3$\cite{Pestka2025,Cao2025} ($\sim$ 5.3 eV) compared to graphene naturally hole-dopes the graphene layer so that the graphene layer supports high-quality 2D plasmon modes without electronic gating\cite{Rizzo2020,Vitalone2024,Rizzo2025}. This design substantially reduces the experimental complexity. 

The interfacial charge transfer from the graphene layer to the surface layers of NiPS$_3$ and the resultant charge-transfer plasmon is confirmed both theoretically by density functional theory (DFT) calculations and experimentally by scattering-type scanning near-field optical microscopy (s-SNOM) measurements. Figure \ref{fig:Fig1}(c) demonstrates the calculated band structure of a graphene-NiPS$_3$ heterostructure with a 3-layer hBN spacer in between. We can clearly observe that the Dirac cone lies above the Fermi level, suggesting a hole-doped graphene layer. Meanwhile, the Fermi level crosses the bottom of the conduction band of NiPS$_3$, indicating light electron doping in NiPS$_3$ beneath the graphene. In order to experimentally confirm the charge-transfer plasmon, we performed the mid-infrared (MIR) s-SNOM measurement with a quantum cascade laser with a wavelength tuned to 11.5~$\upmu$m. Similar to previous observations of charge-transfer plasmons\cite{Rizzo2020,Vitalone2024,Rizzo2025}, we observed plasmonic fringes reflected from the physical graphene edge, as shown in Fig.~\ref{fig:Fig1}(e). The measured mid-infrared plasmonic wavelength is $\lambda_p$ = 233.8 nm, corresponding to a Fermi level at $E_F \sim$ -0.6 eV given the effective dielectric environment of the device, which agrees well with the calculated doping level in the DFT calculation (see Supplemental Information Sec.~VI).

Having the charge transfer plasmon confirmed, we move on to search for its interaction with the antiferromagnetic order and potentially magnons. To match the energy scale of interest, we switch the light source of the s-SNOM measurement to broadband terahertz radiation. For the s-SNOM measurement with THz source (THz-SNOM), we adopt a forward scattering geometry, as indicated in Fig.~\ref{fig:Fig2}(a). An obvious advantage of THz-SNOM is the capability of THz imaging with spatial resolution several orders of magnitude beyond the diffraction limit. The near-field THz image of the same graphene/NiPS$_3$ stack characterized above is shown in Fig.~\ref{fig:Fig2}(b). Under THz illumination, graphene manifests a stark contrast compared to the underlying NiPS$_3$. Upon further analysis by examining the amplitude line-profile across the graphene edge, as demonstrated in Fig.~\ref{fig:Fig2}(c), we could identify a dip-hump structure. This feature serves as the static-state evidence of THz plasmon-polariton, as unveiled by previous THz-SNOM study on the bespoke graphene cavity on top of RuCl$_3$\cite{Vitalone2024}.

Compared to the case in the MIR regime (Fig.~\ref{fig:Fig1}(e)),  extracting the energy–momentum dispersion of plasmon polaritons is less straightforward at THz frequencies, due to the much longer wavelength, which imposes more stringent requirements on sample geometry\cite{Vitalone2024}. However, leveraging the sub-cycle time-resolved information provided by THz pulses, we can bypass the restriction and extract the plasmon-polariton dispersion with a time-domain approach. The recently developed nano-THz spacetime metrology\cite{Xu2024,Xu2025,Xu2026,Anglhuber2025,Rizzo2025} utilizes the edge-reflected THz plasmon-polariton wavepackets and traces its propagation in the two-dimensional plasmonic media. By visualizing the spacetime trajectories (world-lines) of plasmonic wavepackets, we extract plasmon velocities and thus plasmon dispersion and Drude weights across the antiferromagnetic transition of the underlying NiPS$_3$.

A representative space-time map is shown in Fig.~\ref{fig:Fig2}(d). The space-time map is acquired by repeatedly scanning along the aqua dashed line indicated in Fig.~\ref{fig:Fig2}(b) across the graphene edge (black dashed line in both Fig.~\ref{fig:Fig2}(b) and (d)) while varying the THz wave sampling delay time $t$. The raw map consists of both directly tip=scattered THz wave (background) and propagating THz plasmon-polariton wave. Thus, the raw data is then processed by taking partial derivatives along both space ($x$) and time ($t$) axes to remove the spatially and temporally invariant background. More detailed justification of background treatment can be found in previous studies\cite{Xu2024,Xu2025,Xu2026}. The resultant derivative map in Fig.~\ref{fig:Fig2}(d) thus represents the THz plasmon-polariton field signals. From the space-time map, we can unambiguously resolve a group of world-lines. For each point along a world-line at position $x$, the corresponding delay time $t$ is the time when the polaritonic wavepacket accomplishes a round trip from the tip to the edge of graphene and then back to tip. Thus, the group velocity $v_g$ of the wavepacket can be calculated by taking double the reciprocal of the slope of the world-lines ($v_g=2\Delta{x}/\Delta{t}$). In our sample geometry where the graphene is spatially close to the metallic substrate, the plasmon-polariton dispersion is screened to be linear, which can also be verified from the $r_p$ calculation and the similar slopes measured across multiple world-lines. Thus group and phase velocities are identical, $v = v_g = v_{ph}$\cite{Anglhuber2025}

\begin{figure*}[htb!]
\centering
\includegraphics[width=\textwidth]{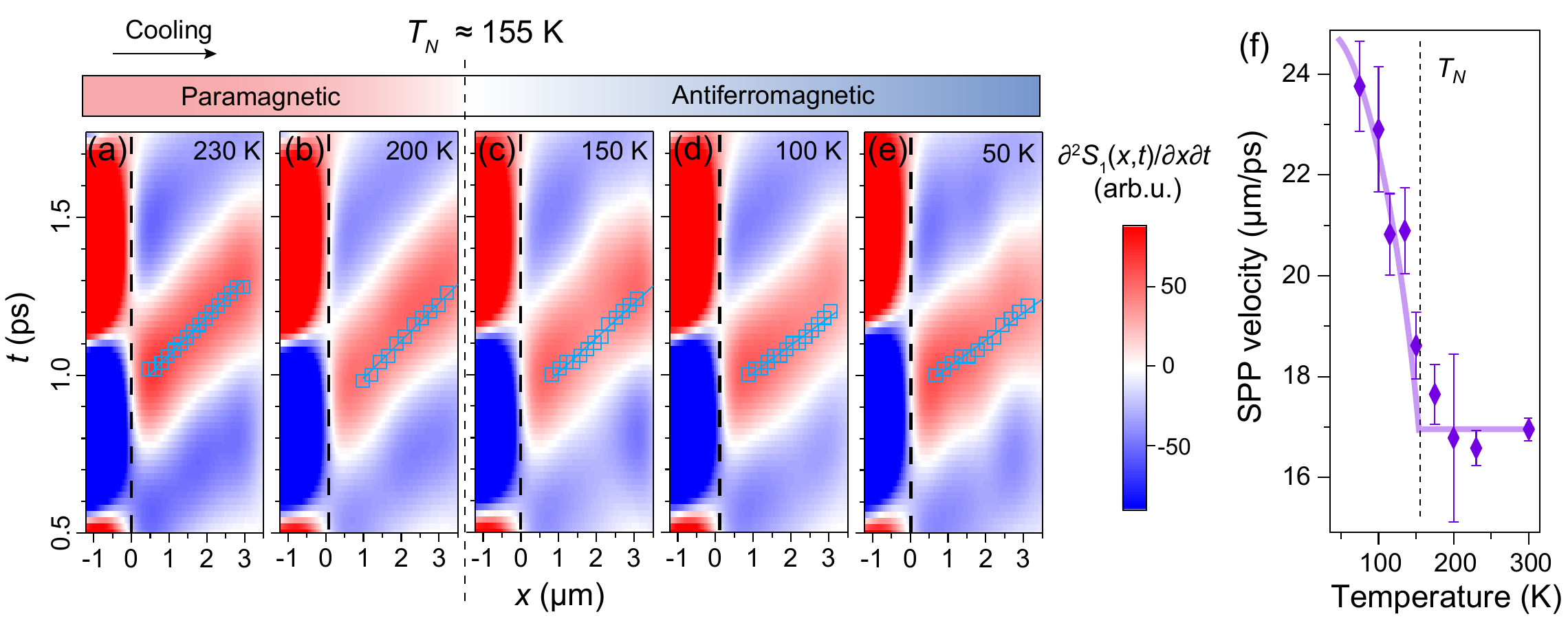}
\caption{\textbf{Tuning of SPP propagation across antiferromagnetic transition in NiPS$_3$.} (a-e)~Spacetime maps of graphene on NiPS$_3$ with hBN spacer at 230 K (a), 200 K (b), 150~K (c), 100 K (d), and 50 K (e). In each panel, a plasmonic world-line that signifies the propagation of SPPs is extracted by highlighting the local maxima of the THz electric field. (f) The group velocity of SPPs as a function of temperature. The group velocities are calculated by doubling the inverse slope of world-lines extracted from spacetime maps. When cooled across the N\'{e}el temperature of NiPS$_3$ ($T_N \approx$ 155 K), the velocity of SPPs starts to increase, nearly scaling with the order parameter of the antiferromagnetic order. The error bars are acquired by calculating the standard deviation across multiple world-lines.}
\label{fig:Fig3}
\end{figure*}

In order to study the potential coupling between the THz plasmon-polariton and the antiferromagnetic order in the underlying NiPS$_3$, we investigate the temperature dependence of the plasmon-polariton velocity by repeating the space-time mapping measurements at various cryogenic temperatures. Figure~\ref{fig:Fig3}(a-e) showcases the space-time maps at various temperatures, cooling across the N\'{e}el temperature ($T_N\approx 155$~K). By fitting the world-lines and extracting the plasmon-polariton velocities, we find a striking upturning behavior in the plasmon-polariton velocities right at $T_N$, as shown in Fig.~\ref{fig:Fig3}(f). This surprising observation was reproduced across multiple devices (see Supplemental Information Sec.~III). The observation that the plasmon-polariton velocity in graphene scales approximately with the antiferromagnetic order parameter in NiPS$_3$ is unlikely to be coincidental and indicates a previously unseen coupling between the plasmon-polariton and the magnetic order, or, possibly, antiferromagnetic magnons. 

\section{Discussion}

The velocity of the plasmon-polariton, which is directly associated with the dispersion of the screened plasmon-polariton, may couple to the underlying magnetic order via two distinct pathways. First, a straightforward possibility would be the direct hybridization between the 1.3 THz magnon resonance in NiPS$_3$ and the acoustic plasmon-polariton, analogous to the previously observed polariton modes such as phonon polaritons or exciton polaritons\cite{Basov2016,Basov2021,Basov2025,Fei2012,Menard2014}. Second, without a magnon-plasmon hybridization gap, the plasmon dispersion could be modified by either a renormalized Drude weight or a modulated dielectric environment. Suppose the dispersion of the screened plasmon-polariton still takes the form of the dispersion of a screened plasmon-polariton
\begin{equation}
    \omega(q) = \sqrt{\frac{dD}{\pi\varepsilon\varepsilon_0}}q
\end{equation}
where $d$ is the distance between the plasmonic media, in our specific case, the graphene, and the metallic substrate (gate), $D$ is the Drude weight, $\varepsilon$ is the environment dielectric function and $\varepsilon_0$ is the vacuum permittivity. The plasmon-polariton velocity could be altered as a result of either Drude weight renormalization or the change in dielectric response across the proximal antiferromagnetic phase transition. 
The coherent hybridization scenario is less likely and can be easily ruled out based on two major reasons: (i) Our THz-SNOM measurement is most sensitive around 0.8-1.0~THz, as a result of the joint contribution from THz source bandwidth and nano-tip enhancement. Thus, suppose such a hybridization takes place, our probe is sensitive to the lower polariton branch and should observe a decrease in plasmon-polariton velocity. This is contradictory to our observation of increased plasmon-polariton velocity in the antiferromagnetic state. (ii) The magnetic resonance in the THz spectrum of NiPS$_3$ is too weak to open a noticeable gap for a NiPS$_3$ thickness of $\sim$100 nm. This can be shown with a Fresnel coefficient calculation ($r_p$) based on the previously measured effective dielectric function of NiPS$_3$\cite{Belvin2021} (see Supplemental Information Sec.~VI). Thus, either the renormalization of the Drude weight or the change in dielectric response of the proximal antiferromagnet would be a more plausible mechanism behind the enhanced plasmon-polariton velocity below $T_N$.

\begin{figure*}[htb!]
\centering
\includegraphics[width=1.01\textwidth]{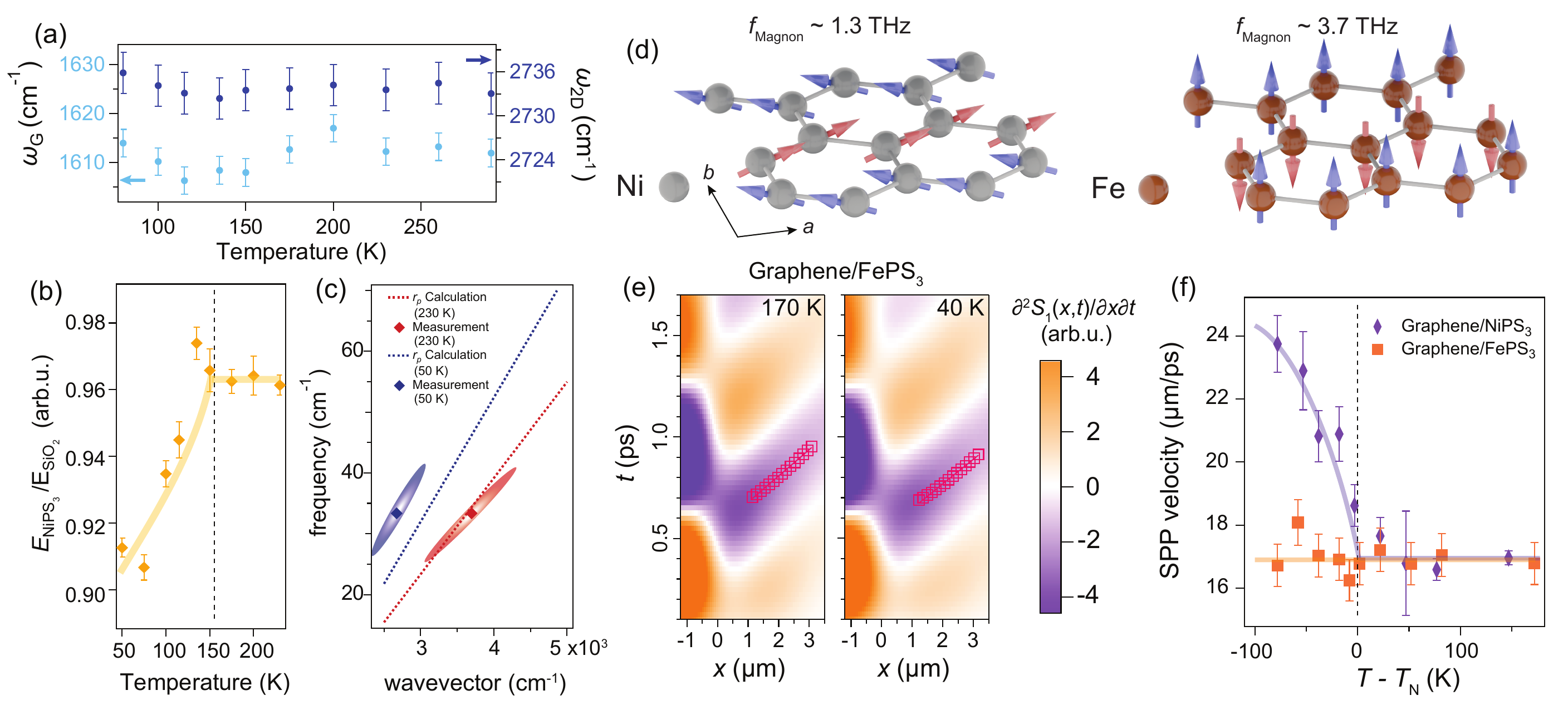}
\caption{\textbf{Possible Mechanisms for Drude weight renormalization.} (a) Phonon frequencies of G mode (left) and 2D mode (right), measured with Raman scattering, as a function of temperature. showing no noticeable shift and thus indicating no change in the carrier density of the graphene layer. (b). The integrated amplitude of reflected electric field on NiPS$_3$ normalized to that of the SiO$_2$ substrate measured with nano-THz time-domain spectroscopy, suggesting a reduction in dielectric constant $\varepsilon_1$ across $T_N$ due to the redistribution of spectral weight by the formation of magnons. (c) Plasmon-polariton dispersion acquired from Fresnel coefficient $r_p$ calculation, above and below $T_N$, with effective dielectric coefficients inferred from respective THz spectra. The red and blue diamond labels the measured value of the spectrum center of mass at 230~K and 50~K, respectively. The length and orientation of the surrounding ellipses denote the slope inferred from plasmon velocity and the spectral range.  (d) Comparison between the magnetic structure of NiPS$_3$ and FePS$_3$. (e) Space-time maps measured on Graphene/FePS$_3$ structure at 170~K (above $T_N\approx118$ K) and 40 K (below $T_N$), showing similar plasmon-polariton velocities. (f) Comparison between the temperature responses of plasmon-polariton velocities in graphene/NiPS$_3$ and graphene/FePS$_3$. Unlike in graphene/NiPS$_3$, the plasmon-polariton velocity shows no noticeable change across $T_N$ in graphene/FePS$_3$.}
\label{fig:Fig4}
\end{figure*}

In the following sections, we analyze the specific mechanisms that could potentially lead to the renormalization of the plasmon-polariton velocity. We first consider carrier density in the graphene layer, which directly affects the Drude weight. 
We thus examine the carrier density in the graphene layer as a function of temperature. For this purpose, we employ Raman scattering to measure the two phonon modes, the G-mode and the 2D-mode, that are known to be a good measure of carrier density in graphene thanks to their coupling to the electronic structure, following the well-established protocol in previous studies\cite{Wang2020,Das2008,Ferrari2013,Heller2016}. As shown in Fig.~\ref{fig:Fig4}(a), the frequencies of both the G mode and 2D mode show no noticeable change across $T_N$, suggesting that there is no sufficiently dramatic increase in charge transfer that induces the increase in plasmon-polariton velocity below $T_N$. This is not particularly surprising, as pronounced temperature-dependent changes in work function—and hence charge transfer—are extremely rare, especially when the electronic band structure stays the same, which is the case in NiPS$_3$ as unveiled by angle-resolved photoemission spectroscopy\cite{Cao2025}. We can thus exclude the contribution from carrier density changes to the Drude weight, and thus the plasmon-polariton dispersion. 

Having ruled out changes in carrier density, a more plausible explanation is a change in the screening to plasmon-polaritons by the underlying environment. To examine the effective dielectric response of the underlying antiferromagnet as a function of temperature, we switched to the third measurement mode of our THz-SNOM instrument: nano-THz time-domain spectroscopy (TDS). We measured the THZ-TDS spectrum on the bare NiPS$_3$ surface, normalizing it with the spectrum on the bare SiO$_2$ to extract the THz reflectance as a function of temperature. At each temperature point, the spectrum is fitted to a constant function, and the resultant fitting values are plotted in Fig.~\ref{fig:Fig4}(b) (See Supplemental Information Sec.~IV for detailed analysis on the spectra).  From the plot, we can clearly observe a decrease in THz reflectance below $T_N$, suggesting a change in the effective dielectric function of NiPS$_3$ in the antiferromagnetic state. As measured previously from far-field THz-TDS, the refractive index is dominated by the real part of the dielectric function $\varepsilon_1$\cite{Belvin2021}, we can approximately assume that the dominant change in the effective dielectric environment is contributed from a reduced effective $\varepsilon_1$ value. We thus calculated the effective dielectric function by combining both nano-THz TDS measurements and the previously reported value\cite{Belvin2021}, followed by a calculation of $r_p$ using the transfer-matrix method\cite{Passler2017}. We then further extracted the plasmon-polariton dispersions at 230~K and 50~K, respectively, as shown in Fig.~\ref{fig:Fig4}(c). We find semi-quantitative agreement between the $r_p$-based calculated plasmon dispersion and the measured velocities. This agreement suggests that the change in the effective dielectric response of the underlying proximal NiPS$_3$ is likely a major source of the renormalized plasmon-polariton velocity in graphene when NiPS$_3$ enters the antiferromagnetic state. We note that the plasmon renormalization should be a general phenomenon when the underlying effective dielectric response is coupled to the antiferromagnetic order, as we can show in a Ginzburg-Landau model (See Sec.~VII in Supplemental Information).

To this end, we would like to take one step further and discuss the microscopic mechanisms behind the changes in the effective dielectric function in NiPS$_3$ when undergoing the antiferromagnetic transition. We notice that a salient feature expected in the THz spectrum that emerges below $T_N$ is the formation of the magnon peak, as observed in previous far-field THz-TDS measurements\cite{Belvin2021}. However, the magnon resonance is not observable in our nano-THz TDS due to the insensitivity to pure magnetic dipoles in the tip-induced scattering process. In this case, there are two possible mechanisms that lead to the change in effective dielectric response. First, the dielectric function could be renormalized due to the multi-magnon continuum as predicted by theory\cite{Little2017,Suzuki1970}. The low-energy two-magnon continuum in NiPS$_3$ was observed in resonant inelastic x-ray scattering (RIXS) measurements\cite {Kang2020}. Alternatively, the reduction in screening could be a result of quenching in magnetic fluctuations when magnetic order forms, as previously shown in a Raman scattering experiment on NiPS$_3$\cite{Kim2019}. One or more of the factors above could reduce the effective screening on the plasmon-polariton in the graphene layer on top. 

To further verify the origin of observed plasmon velocity enhancement and proposed effective dielectric environment renormalization and whether it could be potentially related to magnon formation, we perform a control experiment on a similar heterostructure while replacing NiPS$_3$ with FePS$_3$, another van der Waals antiferromagnetic material in the $M$P$X_3$ ($M$ = Fe, Mn, Ni, ..., $X$ = S, Se) family. Compared to NiPS$_3$, FePS$_3$ also forms a zig-zag antiferromagnetic order along the crystalline $a$-axis below the N\'{e}el temperature ($T_N\approx 118$~K), with the spins aligned, meanwhile, out-of-plane\cite{Zong2023,Zhou2022,Ilyas2024,Park2025}, as indicated in Fig.~\ref{fig:Fig4}(d). Compared to NiPS$_3$, a large Ising anisotropy in FePS$_3$ gives rise to a vastly different magnon frequency (3.7 THz), leaving a clean spectrum in the frequency window where our THz-SNOM measurement is sensitive\cite{Ilyas2024,Lee2023}. Thus, we expect negligible changes in the dielectric function in the frequency range probed by our measurement. Other than the difference in spin orientation and the magnon frequency, FePS$_3$ and NiPS$_3$ are similar in many aspects, such as the crystal structure and chemical composition, and thus should serve as a relatively ideal reference sample for a control experiment. Figure \ref{fig:Fig4}(e) demonstrates the space-time maps for graphene/FePS$_3$ stack at $T>T_N$ (170 K) and $T<T_N$ (40 K). The extracted world-lines exhibit minimal difference in the slope, suggesting the plasmon-polariton enhancement below $T_N$ is absent in graphene/FePS$_3$ interface. This stark difference is better visualized when we plot the plasmon-polariton as a function of temperature next to that of graphene/NiPS$_3$, as shown in Fig.~\ref{fig:Fig4}(f). The absence of plasmon-polariton velocity enhancement supports the potential picture of spectral weight redistribution from the magnon continuum. As further supported by the absence of changes as a function of temperature in nano-THz in FePS$_3$ (See Sec.~IV in Supplemental Information), we can thus conclude that the reduction of effective dielectric screening is a major factor that leads to the plasmon-polariton velocity enhancement. Furthermore, the absence of plasmon renormalization and change in effective dielectric function in FePS$_3$ suggests that quenching of spin fluctuation is less likely to be the mechanism responsible for the dielectric response change in NiPS$_3$ structure. As the Ising anisotropy suppresses fluctuation in FePS$_3$ below $T_N$ while remnant fluctuation is allowed in NiPS$_3$, the spectral redistribution and plasmon renormalization are likely to be more prominent in FePS$_3$ structure if spin fluctuation quenching is predominantly responsible for the phenomena, and this leaves the multi-magnon continuum a more plausible scenario. 

While we conclude from nano-THz TDS and control experiments on graphene/FePS$_3$ that the effective dielectric response of underlying antiferromagnetic order should be accounted for in the plasmon-polariton velocity enhancement, we would like to note that there are still other possibilities that could lead to this phenomenon. Specifically, the graphene Dirac band dispersion, and thus the Fermi velocity, as another major factor that decides the Drude weight, can also possibly contribute to the plasmon-polariton velocity enhancement. This could happen if the graphene band structure is renormalized by the strong correlation in NiPS$_3$\cite{Kang2020}. This possibility is subject to potential future studies with photoemission spectroscopy\cite{Lv2019Angle-resolvedMaterials}.

Though the direct Rabi-splitting type of interaction between antiferromagnetic magnons and plasmons or plasmon-polaritons is theoretically predicted to be achievable\cite{Costa2023,Yuan2025,Gorobets2025,Dyrda2023}, the coupling in realistic material systems, such as the ones we study in this work, may be several orders of magnitude weaker due to that plasmon scattering rate is larger than the coupling strength, thus making such kinds of interaction extremely hard to observe. In order to enhance the magnon-plasmon interaction strength in van der Waals heterostructures, future attempts could potentially be made to explore novel van der Waals antiferromagnetic materials with a larger magnon oscillator strength\cite{Golovchanskiy2021,Xiong2024,Martinez2026} or a layered cavity structure that enhances light-spin interaction\cite{Norcia2018,Fan2026,Zhang2014}. Nevertheless, our observation of plasmon-polariton velocity renormalization in graphene/NiPS$_3$ showcases a potential pathway that allows an indirect interaction between the two quasiparticles that were traditionally regarded as nearly impossible to hybridize. We would also like to note that the redistribution of spectral weight due to the bosonic excitation may have an effect on a wider range of plasmonic heterostructures with a variety of interactions between plasmon-polaritons and other bosonic excitations, subject to further future studies.  

\section{Conclusions}

Utilizing s-SNOM measurements in both mid-infrared and terahertz regimes, assisted by density functional theory calculations, we unambiguously characterized the charge transfer plasmon-polariton in the graphene/NiPS$_3$ heterostructure, a novel prototype plasmonic platform we engineered for realizing magnon-plasmon interaction in quantum materials. The versatile channels in the cryogenic THz-SNOM: imaging, spacetime mapping, and spectroscopy, respectively allowed us to image the charge transfer plasmon-polariton, extract the plasmon-polariton velocity, and characterize the effective dielectric environment, all as a function of temperature. We observed an unconventional enhancement in the plasmon-polariton velocity in graphene when the underlying NiPS$_3$ undergoes the antiferromagnetic transition. Leveraging additional Raman scattering measurements, control experiments on graphene/FePS$_3$ structure, and near-field THz-TDS, we concluded that the plasmon velocity enhancement is likely a result of changes in dielectric screening due to the spectral weight redistribution that could originate from magnon continuum formation. This result marks the cornerstone of a rare experimental observation of interaction between antiferromagnetic order and plasmon-polaritons. The fact that plasmons are sensitive to magnetic order, and thereby implicitly to fluctuations around that order, strongly suggests the possibility of realizing coherent and strong plasmon–magnon coupling in such van der Waals heterostructures. These findings herald a new frontier in both the fundamental physics of collective excitations and the potential applications in the merging fields of plasmonics, nanophotonics, and spintronics.

\section*{Acknowledgments}
The authors thank Carina Belvin, Edoardo Baldini, Clifford Allington, and Nuh Gedik for providing far-field THz time-domain spectroscopy data from a previous publication and for helpful discussion. The authors thank Alfred Zong, Anshul Kogar, Kenneth Burch, and Liuyan Zhao for helpful discussion. 

Measurements at Columbia are supported by ARO-W911NF2510062. Crystal growth and characterization at Columbia were supported by a Brown Investigator Award (CALTECH S681506), a program of the Brown Institute for Basic Sciences at the California Institute of Technology (X.R.). K.W. and T.T. acknowledge support from the CREST (JPMJCR24A5), JST and World Premier International Research Center Initiative (WPI), MEXT, Japan. This work was supported by the Max Planck–New York City Center for Non-Equilibrium Quantum Phenomena and by the Cluster of Excellence Advanced Imaging of Matter (AIM). The Flatiron Institute is a division of the Simons Foundation

Y.S. and S.X. conceived the study. Y.S, S.X., and R.A.V. performed s-SNOM measurements. Y.S., Z.L., J.J., and Y.W. fabricated the sample stacks. D.G.C and V.d.S.L.A synthesized bulk NiPS$_3$ and FePS$_3$ crystals under the supervision of X.R.. T.T. and K.W. synthesized high-quality hBN crystals.  J.C.H., C.R.D., and X.R. advised synthesis and fabrication efforts. E.V.B and N.W. performed density functional theory calculations advised by A.R.. Y.S., Z.Z., and C.-Y.H. performed Raman scattering measurements advised by X.-Y.Z. Y.S. wrote the manuscript with input from all authors. The work was supervised by D.N.B..

\section*{Data Availability}
The data that support the plots within this paper and other findings of this study are available from the corresponding author upon reasonable request.

\section*{Methods}
\subsection{Sample preparation}

\subsubsection*{Crystal growth}
NiPS$_3$ and FePS$_3$ were synthesized using high-purity reagents iron wire (99.995\%, 1mm dia, Thermo Scientific) or nickel powder (99.999\%,  Thermo Fischer), red phosphorus chunks (99.99\%, Aldrich), and sulfur pieces (99.9995\%, Alfa Aesar), used as received. Stoichiometric amounts of the reagents, with a 5\% excess of S to account for its higher vapor pressure, were weighed to 1g total mass and loaded into 12.7-mm o.d., 10.5-mm i.d. fused silica tubes. The atmosphere in the tubes was changed to an Ar atmosphere and 20~mg of I$_2$ was added to act as a transport agent. The tubes were then sealed under low pressure while cooled with liquid nitrogen to 12~cm length.

The NiPS$_3$ tube was subjected to the following heating profile using a two-zone computer-controlled horizontal tube furnace: Source side: Heat to 500~$^\circ$C in 24 hours, soak for 12 hours, heat to 750~$^\circ$C in 6 hours, soak for a week, and then cool to room temperature in 6 hours. Sink side: Heat to 550~$^\circ$C in 24 hours, soak for 12 hours, heat to 700~$^\circ$C in 6 hours, soak for a week, and then cool to room temperature in 6 hours.

The FePS$_3$ tube was subjected to the following heating profile using a two-zone computer-controlled horizontal tube furnace: Source side: Heat to 575~$^\circ$C in 24 hours, soak for 4 days, heat to 710~$^\circ$C in 6 hours, soak for 5 days, and then cool to room temperature in 6 hours. Sink side: Heat to 525~$^\circ$C in 24 hours, soak for 4 days, heat to 660~$^\circ$C in 6 hours, soak for 5 days, and then cool to room temperature in 6 hours. FePS$_3$ was subject to longer pre-reaction times and with a temperature gradient, so that the formed crystals would not passivate the Fe wire surface. After cooling, the tubes were opened and washed with multiple washings of acetonitrile, to remove the I$_2$, and toluene, to remove excess sulfur.

\subsubsection*{Device fabrication}

Thin flakes of graphene and NiPS$_3$/FePS$_3$ are mechanically exfoliated onto plasma-cleaned Si/SiO$_2$ substrates with a 285~nm oxide layer. The thickness and uniformity of the exfoliated flakes are identified and characterized using optical microscopy. The heterostructure of thin flakes is assembled using the van der Waals dry transfer technique. A top BN flake is first picked up by polycarbonate (PC) transfer polymer rested on a polydimethylsiloxane (PDMS) stamp. The top hBN flake is then used to pick up the graphene flake and a thin hBN spacer layer if needed. Finally, the stack is released from the polymer stamp onto the exfoliated NiPS$_3$/FePS$_3$ flakes at 180$^\circ$~C.

\subsection{Mid-infrared scattering-type Scanning Near-field Optical Microscopy (s-SNOM)}
The MIR s-SNOM measurements were performed on a commercial Neaspec system under ambient conditions using commercial ArrowTM AFM probes with nominal resonant frequencies of $f$ = 75 kHz. Tunable continuous wave quantum cascade lasers produced by Daylight Solutions were used, spanning wavelengths from 9 to 11.7 μm. The detected signal was demodulated at the fourth harmonic of the tip tapping frequency in order to minimize farfield contributions to the scattered light. The fourth harmonic of the near-field scattering amplitude ($S4$) is collected using a pseudoheterodyne interferometry technique.

\subsection{Terahertz s-SNOM and spacetime mapping}
The THz s-SNOM measurements were performed in a home-built cryogenic system under an ultrahigh vacuum condition using AFM tips produced by Rocky Mountain Nanotechnology, LLC with a nominal resonant frequency of 30 – 80 kHz. The THz broadband pulse is generated and detected using a pair of photo-conductive antennas (PCAs, Menlo Systems GmbH). The THz radiation from the PCA emitter is collimated by a TPX lens and focused onto the tip and sample by a parabolic mirror. The scattered field is detected by an unbiased PCA in the time domain using a 50 fs, 780 nm gate beam, and the photocurrent signal is demodulated by a lock-in amplifier at harmonics of the tip tapping frequency.\\

\subsection{Raman scattering}
The Raman scattering measurements were carried out on a home-built setup with a Nikon TE-300 inverted microscope. A 633 nm HeNe laser first passed a linear polarizer and was filtered by a reflective band-rejection filter (Ondax), then sent through a half-wave plate to control the angle of the incident polarization, and into the microscope objective (40x, NA = 0.6). The incident light was focused on the sample mounted in a cryostat supplied with liquid nitrogen. The Raman scattered light was collected through the same objective and sent through a long-pass filter (RazorEdge) for filtering out the Rayleigh scattering light. The Raman signal was subsequently focused onto the entrance slit of a spectrometer (Princeton Instruments HRS-300) with a 600 gr/mm grating that dispersed the spectrum onto a liquid-nitrogen-cooled CCD camera (Princeton Instruments LN400/B). 

\subsection{First principles parameterization}
To obtain the bandstructure and charge transfer of the heterostructure, we performed density functional theory (DFT) calculations. The simulations were performed by first calculating the charge density and electric potential of each subsystems separately (i.e., $\mathrm{NiPS_3}$, $h$BN and graphene), and then subtracting these quantities from those of the full heterostructure to obtain the density difference~\cite{Bokdam2011}
\begin{align}
 \Delta n &= n_{{\rm NiPS}_3-h{\rm BN}-{\rm Gr}} - n_{{\rm NiPS}_3} - n_{h{\rm BN}} - n_{\rm Gr}
\end{align}
In all calculations, we used a heterostructure with three $h$BN layers between single layers of $\mathrm{NiPS_3}$ and graphene.

The DFT calculations were performed with the {\sc Octopus} electronic structure code~\cite{TancogneDejean2017,TancogneDejean2020}, within the local density approximation, using norm conserving pseudopotentials from the Pseudodojo project. To treat the local correlations on the Ni atoms, we used the hybrid DFT+$U$ functional ACBN0~\cite{TancogneDejean2017}, with a value of $U = 2$ eV. The calculations were performed in a $2 \times 2$ hexagonal $\mathrm{NiPS_3}$ supercell, with a length of $12.3$ {\AA} along each lattice vector, corresponding to a $5 \times 5$ supercell for $h$BN and graphene. We used mixed boundary conditions, periodic in the in-plane directions and open in the out-of-plane direction, with a vacuum region of $35$ {\AA} to ensure convergence in the out-of-plane direction. A $3\times 3$ $k$-point grid and a real-space grid spacing of $0.3$ Bohr were employed.

\end{document}